\documentclass[11pt]{article}

\usepackage[dvips]{graphicx}
\usepackage{psfrag}
\usepackage{amssymb}
\usepackage{amsmath}
\usepackage{multirow}

\newcommand{\bea}{\begin{eqnarray}}
\newcommand{\eea}{\end{eqnarray}}
\newcommand{\beq}{\begin{equation}}
\newcommand{\eeq}{\end{equation}}
\newcommand{\ec}{\end{center}}
\newcommand{\bc}{\begin{center}}

\newcommand{\pdir}{p\kern -5.2pt\raise 0.2ex\hbox {/}}

\newcommand{\vdir}{v\kern -5.75pt\raise 0.15ex\hbox {/}}
\newcommand{\kdir}{k\kern -5.75pt\raise 0.15ex\hbox {/}}
\newcommand{\epsdir}{\epsilon\kern -5.0pt\raise 0.15ex\hbox {/}}
\newcommand{\Ddir}{D\kern -7.75pt\raise 0.20ex\hbox {/}}
\newcommand{\Adir}{A\kern -7.75pt\raise 0.20ex\hbox {/}}
\newcommand{\ldir}{l\kern -5.0pt\raise 0.2ex\hbox{/}}
\newcommand{\varepsdir}{\varepsilon\kern -5.5pt\raise 0.15ex\hbox{/}}

\def\at2{A_T^{(2)}(q^2)}
\newcommand{\atre}{A_T^{({\rm re})}(q^2)}
\newcommand{\atim}{A_T^{({\rm im})}(q^2)}

\newcommand{\eqa}[1]{\begin{eqnarray} #1 \end{eqnarray}}
\newcommand{\al}[1]{\begin{align} #1 \end{align}}

\newcommand{\nn}{\nonumber}

\newcommand{\azeL}{{A_0^L}}
\newcommand{\azeR}{{A_0^R}}

\newcommand{\apeL}{{A_\perp^L}}
\newcommand{\apeR}{{A_\perp^R}}

\newcommand{\apaL}{{A_\|^L}}
\newcommand{\apaR}{{A_\|^R}}

\newcommand{\re}{{\rm Re}}

\begin{document}


\begin{flushright}
{\small
UAB-FT 722\\
}
\end{flushright}
$\ $
\vspace{1.5cm}
\begin{center}
\Large\bf
On
the S-wave pollution of $B\to K^* l^+l^-$ observables
\end{center}

\vspace{3mm}
%

\begin{center}
{\sc  Joaquim Matias}\\[2mm]
{\em 
Universitat Aut\`onoma de Barcelona, 08193 Bellaterra, Barcelona, Spain
}
\end{center}

\vspace{1mm}
\begin{abstract}\noindent
It has been argued recently that transverse asymmetries  that are expected to be shielded from the presence of the S-wave $K\pi$ pairs originating from the decay of a scalar $K_0^*$ meson, are indeed affected by this pollution due to the impossibility to extract cleanly the normalization for these observables. In this short note we show how using folded distributions, which is nowadays the preferred method to obtain the information from the 4-body decay mode $B \to K^*(\to K\pi) l^+l^-$, one can easily bypass this problem and extract the clean observables $P_{1,2,3}$ and also $P_{4,5,6}^\prime$  in a way completely free from this pollution including all lepton mass corrections. We also show that in case one insists in using uniangular distributions  to extract these observables it is possible to reduce this pollution to just lepton mass suppressed terms.   
On the contrary, the $S_i$ observables, that are by definition normalized by the full differential decay distribution, will indeed suffer from this pollution via their normalization.
Finally, we also present a procedure to minimize the error associated to neglecting lepton mass corrections in the distribution defining a massless-improved limit. 
\end{abstract}


\newpage
The angular distribution $B \to K^*(\to K \pi)l^+l^-$ is nowadays, but also in the near future, one of the central players in our search for New Physics in rare B decays. A wide variety of  experimental analysis have been presented~\cite{exp1,exp2,exp3,exp4} focusing on different observables related to this rich mode. Simultaneously, there has been a huge effort from the theory community 
to provide a complete  and accurate description of this 4-body decay mode \cite{th1}-\cite{damir}.
Given the relevance of this mode in providing new constraints \cite{cons1}-\cite{DescotesGenon:2012zf} it is of utmost importance to control and isolate any possible source of hadronic pollution that can spoil a clear signal of the theory that lies beyond the SM. 

In a recent  interesting paper \cite{damir} one  possible source
of pollution of the angular distribution $B \to K^*(\to K \pi)l^+l^-$ due to events coming from the distribution $B\to K_0^*(\to K \pi) l^+l^-$, where $K_0^*$ is a scalar meson resonance, was analyzed. There it was presented the full distribution of the combined channels $B \to K^*(\to K \pi) l^+l^-$ and $B \to K_0^*(\to K \pi) l^+l^-$ and the focus was on the impact that this pollution could have on the extraction of the transverse asymmetries $A_T^{(2)}$ \cite{me}, $A_T^{({\rm re})}$\cite{th8} and $A_T^{({\rm im})}$\cite{th8}.
In \cite{damir} it was argued that even if one should expect that the transverse asymmetries are unaffected by this S-wave contamination, they are afflicted from this disease via their normalization. They conclude that if the transverse asymmetries (defined with an unusual normalization)  are measured in the particular way they propose, they are afflicted by this pollution at the level of less than 10$\%$ in all $q^2$-range  except for $q^2=2$ GeV$^2$ where the pollution can be as large as $23 \%$. 

In this short note we  present a simple procedure to bypass this pollution and extract the transverse asymmetries defined as in \cite{primary} in a way completely free from any S-wave contamination  including also all lepton mass corrections. First, we  show  using folded distributions, that an exact ($m_\ell \neq 0$) and clean (no S-wave pollution)  extraction of all the $P_i$ observables \cite{primary} is perfectly possible.  Second, we argue that even using uniangular distributions a strategy can be designed to reduce the impact of the companion decay on the transverse asymmetries to only lepton mass suppressed terms.  Incidentally, we also present two  procedures to introduce lepton mass corrections inside the observables in the distribution, an exact one preserving the structure of the massless case as much as possible and an approximated one that minimizes the error of neglecting lepton masses.  
The  results presented here aim at reducing substantially two main sources of systematic errors of the experimental data namely the  S-wave pollution and lepton mass corrections. The final goal is to provide strategies to enhance the sensitivity to any possible signal of New Physics affecting the 
 clean observables $P_{1,2,3}$, $P_{4,5,6}^\prime$.

In Section I we  recall the definitions of the known observables, introduce new ones and we  also present a so called ``massless-improved limit". In Section II we show how to extract the $P_i$ observables in an exact way using folded distributions including all lepton mass corrections with zero S-wave pollution. Also a less clean approximate procedure  is discussed using uniangular distributions. In  appendices A and B we define the coefficients of the distribution coming from the companion scalar decay  $B\to K_0^*(\to K \pi) l^+l^-$ and provide more examples of folded distributions.
\section*{\normalsize I. Definitions: Observables, full distribution and massless-improved limit}
 
The angular distribution that describes the four-body decay $B\to K^*(\to K\pi) l^+l^-$ including  the S-wave pollution from the companion decay $B\to K_0^*(\to K\pi) l^+l^-$ is \cite{me,th4,damir}
 \eqa{\label{dist}
\frac{d^4\Gamma}{dq^2\,d\!\cos\theta_K\,d\!\cos\theta_l\,d\phi}&=&\frac9{32\pi} \bigg[
J_{1s} \sin^2\theta_K + J_{1c} \cos^2\theta_K + (J_{2s} \sin^2\theta_K + J_{2c} \cos^2\theta_K) \cos 2\theta_l\nn\\[1.5mm]
&&\hspace{-2.7cm}+ J_3 \sin^2\theta_K \sin^2\theta_l \cos 2\phi + J_4 \sin 2\theta_K \sin 2\theta_l \cos\phi  + J_5 \sin 2\theta_K \sin\theta_l \cos\phi \nn\\[1.5mm]
&&\hspace{-2.7cm}+ (J_{6s} \sin^2\theta_K +  {J_{6c} \cos^2\theta_K})  \cos\theta_l    
+ J_7 \sin 2\theta_K \sin\theta_l \sin\phi  + J_8 \sin 2\theta_K \sin 2\theta_l \sin\phi \nn\\[1.5mm]
&&\hspace{-2.7cm}+ J_9 \sin^2\theta_K \sin^2\theta_l \sin 2\phi \bigg]\,X+ S
} 
where
\eqa{
S&=&\frac{1}{4\pi} \left[ {\tilde J}_{1a}^c+{\tilde J}_{1b}^c \cos \theta_K+ ({\tilde J}_{2a}^c+{\tilde J}_{2b}^c \cos\theta_K ) \cos 2\theta_\ell + {\tilde J}_{4} \sin \theta_K \sin 2 \theta_\ell \cos \phi \right.\nn\\ &&\left.+
{\tilde J}_{5} \sin \theta_K \sin \theta_\ell \cos \phi + {\tilde J}_{7} \sin \theta_K \sin \theta_\ell \sin \phi+
{\tilde J}_{8} \sin \theta_K \sin 2\theta_\ell \sin \phi\right]
}
and
\eqa{X=\int dm_{K\pi}^2 |BW_{K^*}(m_{K\pi}^2)|^2}  being a correction introduced in \cite{damir} to take into account the width of the resonance (see \cite{damir} for precise definition and details on this  function). This correction factorizes from the standard $J_i$ defined in \cite{me,th2,th4} and as, we will show, it will always cancel in the $P_i$ observables. The ${\tilde J}_i$ containing the interference terms  coming from the $B \to K^*_0(\to K\pi) l^+l^-$ decay   computed in \cite{damir} are defined in Appendix A.

The full differential decay distribution is then
\eqa{\frac{d\Gamma_{full}}{dq^2}=\frac{d\Gamma_{K^*}}{dq^2}+\frac{d\Gamma_{K_0^*}}{dq^2}
}
with 
\eqa{ \label{full}
\frac{d\Gamma_{K^*}}{dq^2}=\frac{1}{4}\left(3 J_{1c} + 6 J_{1s}-J_{2c}-2 J_{2s} \right) X, \quad \frac{d\Gamma_{K_0^*}}{dq^2}=+ 2 {\tilde J}_{1a}^c - \frac{2}{3} {\tilde J}_{2a}^c
}

 Here we will not consider scalar contributions to facilitate the comparison with \cite{damir} where they were also neglected. Notice that the distribution including lepton masses (but neglecting scalars $J_{6}^c=0$) contains 11 $J_i$ coefficients (only 10 are independent \cite{matias2,primary}) plus 8 ${\tilde J}_i$ coefficients coming from the interference with the S-wave polluting process.  A generalization of the results presented here to include scalars  is straightforward. We take the same inputs as in \cite{DescotesGenon:2012zf} and all observables should be understood as $J \to J+{\bar J}$ (obviously the same applies for expressions written in terms of transversity amplitudes).

The first important point is the exact definition of the transverse asymmetries in terms of the coefficients of the distribution. In \cite{damir} the transverse asymmetries are defined with an unusual normalization
\begin{align}\label{eq:at}
{{\tilde A}_T^{(2)}} = {4 J_3(q^2)\over 3 J_{1s}(q^2)- J_{2s} (q^2)}&\,,\quad \atim = {4 J_9(q^2)\over 3 J_{1s}(q^2)- J_{2s} (q^2)}\,,\nn\\
 \atre = &{\beta_\ell J_{6s}(q^2)\over 3 J_{1s}(q^2)- J_{2s} (q^2)}\,.\quad 
\end{align}
This choice of normalization in \cite{damir} is driven by the aim of extracting it from 
 the $\theta_K$ uniangular distribution (see below). These definitions albeit correct have two main disadvantages: 
 \begin{itemize} 
 \item  The expression of the transverse asymmetries given in  Eq.(\ref{eq:at}) once expressed in terms of transversity amplitudes involves lepton mass terms via the coefficient\footnote{From now on we will distinguish between  two types of lepton mass dependences:  lepton mass terms that will refer to terms like the last term in Eq.(\ref{j1s}) and mass dependence via the prefactor $\beta_\ell$.}
\eqa{\label{j1s} J_{1s}  & = & \frac{(2+\beta_\ell^2)}{4} \left[|\apeL|^2 + |\apaL|^2 +|\apeR|^2 + |\apaR|^2 \right]
    + \frac{4 m_\ell^2}{q^2} \re\left(\apeL\apeR^* + \apaL\apaR^*\right)\,}
where $\beta_\ell=\sqrt{1-4 m_\ell^2/q^2}$. This implies that their definition is not invariant under changes of the lepton mass ($\ell=e,\mu$).
\item Normalization requires to measure two  coefficients of the distribution.
\end{itemize}

 For these two reasons we prefer to stick to the definition of the $P_i$ observables as given in \cite{primary}. These definitions in terms of transversity amplitudes  do not  dependent   on any  lepton mass term:
  \eqa{
P_1&=&\frac{|A_\perp^L|^2-|A_\|^L|^2+(L\leftrightarrow R)}{|A_\perp^L|^2+|A_\|^L|^2+(L\leftrightarrow R)}=\frac{ J_{3}}{2 J_{2s}}\ ,\label{p1}\ \nn \\
P_2&=&\frac{{\rm Re}(A_\perp^{L*} A_\|^L-A_\perp^R A_\|^{R*})}{|A_\perp^L|^2+|A_\|^L|^2+(L\leftrightarrow R)}=\beta_\ell\frac{J_{6s}}{8 J_{2s}}\ ,\nn \ \quad \\
P_3&=&\frac{{\rm Im}(A_\perp^{L*} A_\|^L-A_\perp^R A_\|^{R*})}{|A_\perp^L|^2+|A_\|^L|^2+(L\leftrightarrow R)}=-\frac{J_9}{4 J_{2s}}
\label{props}}
These observables were completed in \cite{DescotesGenon:2012zf} with 
\eqa{P_4^\prime&=&\frac{\sqrt{2}{\rm Re}(A_0^{L*} A_\|^L+A_0^R A_\|^{R*})}{\sqrt{\left(|A_\perp^L|^2+|A_\|^L|^2+(L\leftrightarrow R)\right)\left(|A_0^L|^2+(L \leftrightarrow R)\right)}}=\frac{J_{4}}{\sqrt{-J_{2c} J_{2s}}}\ ,\nn \ \quad \\
P_5^\prime&=&\frac{\sqrt{2}{\rm Re}(A_0^{L*} A_\perp^L-A_0^R A_\perp^{R*})}{\sqrt{\left(|A_\perp^L|^2+|A_\|^L|^2+(L\leftrightarrow R)\right)\left(|A_0^L|^2+(L \leftrightarrow R)\right)}}=\frac{\beta_\ell}{2}\frac{J_{5}}{\sqrt{-J_{2c} J_{2s}}}\ ,\nn \ \quad \\
P_6^\prime&=&\frac{\sqrt{2}{\rm Im}(A_0^{L*} A_\|^L+A_0^R A_\|^{R*})}{\sqrt{\left(|A_\perp^L|^2+|A_\|^L|^2+(L\leftrightarrow R)\right)\left(|A_0^L|^2+(L \leftrightarrow R)\right)}}=-\frac{\beta_\ell}{2}\frac{J_{7}}{\sqrt{-J_{2c} J_{2s}}}\ 
\label{props2}}
where the $\beta_\ell$ prefactor included in $P_{2,5',6'}$ is there to ensure that the expression in terms of transversity amplitudes does not contain any lepton mass dependence. $P_1$ is the transverse asymmetry $A_T^{(2)}$ proposed in \cite{me} and $P_2$ and $P_3$ correspond to the $A_T^{({\rm re})}$ and $A_T^{({\rm im})}$, respectively, normalized as in \cite{th8} and not like in Eq.(\ref{eq:at}).
Notice also that the normalization of the transverse asymmetries in  Eqs.(\ref{props}) requires to measure one single coefficient of the distribution, namely, $J_{2s}$, free from any lepton mass term. 


It is very convenient  to express the coefficients of the angular distribution directly in terms of observables. 
In  Ref.~\cite{primary} it was presented the most complete expression of the coefficients of the distribution  including lepton masses and scalars  in terms of the full basis of  observables. Later on, in \cite{DescotesGenon:2012zf} a very compact parametrization was found  for the massless case without scalar contributions. Here we present a  generalization of Ref. \cite{DescotesGenon:2012zf}  to include lepton masses with a similar structure to the massless case  at the price of  trading the original basis \cite{primary}  $$\Big\{ \frac{d\Gamma_{K^*}}{dq^2}, A_{\rm FB}, P_{1,2,3}, P_{4,5,6}, M_1, M_2\Big\} \to \Big\{{\hat F}_T \frac{d\Gamma_{K^*}}{dq^2},{\hat F}_L \frac{d\Gamma_{K^*}}{dq^2}, {\tilde F}_T \frac{d\Gamma_{K^*}}{dq^2}, {\tilde F}_L\frac{d\Gamma_{K^*}}{dq^2}, P_{1,2,3},P_{4,5,6}^\prime  \Big\}$$ The exact parametrization of the coefficients in the massive case is then\footnote{Notice that in the first coefficient $J_{1s}$ the $\beta_\ell$ dependence does not appear explicitly (but it is included inside ${\hat F}_T$). The reason is that parametrized in this way and only in this coefficient the mass dependence piece of $\beta_\ell$ will help later on to reduce the contribution from the lepton mass term.}  
 \al{\label{Jsintermsobs}
J_{1s} &=   \dfrac{3}{4} {\hat F}_T \frac{1}{X}\frac{d\Gamma_{K^*}}{dq^2}  , &
& J_{2s} =  \dfrac{1}{4} \beta_\ell^2 {\tilde F}_T \frac{1}{X}\frac{d\Gamma_{K^*}}{dq^2},    \nn \\
J_{1c} &= {\hat F}_L \frac{1}{X}\frac{d\Gamma_{K^*}}{dq^2} \ ,&
&J_{2c} = -\beta_\ell^2 {\tilde F}_L \frac{1}{X}\frac{d\Gamma_{K^*}}{dq^2} \ ,\nn\\
J_3  &= \dfrac{1}{2} \beta_\ell^2 P_1 {\tilde F}_T \frac{1}{X}\frac{d\Gamma_{K^*}}{dq^2}  ,&
& J_{6s} = 2 \beta_\ell P_2  {\tilde F}_T \frac{1}{X}\frac{d\Gamma_{K^*}}{dq^2}  ,\nn\\
J_4 &= \dfrac1{2}\beta_\ell^2 {P_4'}  \sqrt{{\tilde F}_T {\tilde F}_L}\ \frac{1}{X}\frac{d\Gamma_{K^*}}{dq^2} 
 ,&
& J_9 = - \beta_\ell^2 P_3  {\tilde F}_T \frac{1}{X}\frac{d\Gamma_{K^*}}{dq^2} \ ,\nn\\
 J_5 &= \beta_\ell P_5' \sqrt{{\tilde F}_T {\tilde F}_L}\ \frac{1}{X}\frac{d\Gamma_{K^*}}{dq^2}  \ ,\nn\\
J_7 &= -\beta_\ell P_6' \sqrt{{\tilde F}_T {\tilde F}_L }\ \frac{1}{X}\frac{d\Gamma_{K^*}}{dq^2} \ ,   }
and the redundant but interesting coefficient $J_8$ is
\eqa{\label{redundant}J_8=-\frac{1}{2} \beta_\ell^2 Q^\prime\sqrt{{\tilde F}_T {\tilde F}_L }\ \frac{1}{X}\frac{d\Gamma_{K^*}}{dq^2}} 
where  the new observables ${\hat F}_{T,L}$, ${\tilde F}_{T,L}$ are given by
\bea {\hat F}_T&=& \frac{16}{3 N} \left[ \frac{ (2+\beta_\ell^2)}{4} \left[|\apeL|^2 + |\apaL|^2 +|\apeR|^2 + |\apaR|^2 \right]
    + \frac{4 m_\ell^2}{q^2} \re\left(\apeL\apeR^* + \apaL\apaR^*\right)\right]=\frac{16 J_{1s}}{3N} \nn \\ 
   {\hat F}_L&=& \frac{4}{N}  \left[|\azeL|^2 +|\azeR|^2  + \frac{4m_\ell^2}{q^2} \left[|A_t|^2 + 2\re(\azeL^{}\azeR^*) \right] \right] =\frac{4 J_{1c}}{N}  
  \nn \\
  {\tilde F}_T&=&\frac{4}{N} \left[ |\apeL|^2+ |\apaL|^2 + |\apeR|^2+ |\apaR|^2\right]=\frac{16 J_{2s}}{\beta_\ell^2 N} \nn \\
  {\tilde F}_L&=&\frac{4}{N} \left[|\azeL|^2 + |\azeR|^2 \right]=-\frac{4 J_{2c}}{\beta_\ell^2 N}  
    \eea
  and the normalization is
   \bea 
 N&=& \left[(3+\beta_\ell^2) \left( |\apeL|^2+ |\apaL|^2  + |\azeL|^2 + |\apeR|^2+ |\apaR|^2  + |\azeR|^2 \right) +\right. \nn \\ &&\left.+  \frac{12m_\ell^2}{q^2} \left[|A_t|^2 + 2\re(\azeL^{}\azeR^* + \apeL\apeR^* + \apaL\apaR^*  ) \right]     \right] =  3 J_{1c} + 6 J_{1s}- J_{2c}- 2 J_{2s} =\frac{4}{X} \frac{d \Gamma_{K^*}}{dq^2}. \nonumber
      \eea
At this point several remarks are in order:
\begin{itemize}
\item[I.] $P_i$ and $P_j^\prime$ $q^2$-dependent observables once expressed in terms of transversity amplitudes do not dependent on $m_\ell$, on the contrary ${\hat F}_{T,L}$ and ${\tilde F}_{T,L}$ do depend.
\item[II.] Eqs.(\ref{Jsintermsobs}) and Eq.(\ref{redundant}) are exact concerning lepton masses (no scalars included). In all the folded distributions discussed in next section we will always use this exact parametrization.
\item[III.] They are constructed to resemble Eqs.(14) of \cite{DescotesGenon:2012zf} for the massless case. The main difference with the massless case is that here there are two longitudinal polarization fractions and two transverse polarization fractions and the sums of transverse and longitudinal fractions differ slightly from one:
$$ {\hat F}_L+{\hat F}_T=1 + {\hat \delta} \quad \quad {\tilde F}_T+{\tilde F}_L=1 + \delta $$ 
In the exact massless limit ${\hat F}_{L}|_{m_\ell=0}={\tilde F}_{L}|_{m_\ell=0}$, ${\hat F}_{T}|_{m_\ell=0}={\tilde F}_{T}|_{m_\ell=0}$ and ${\hat \delta}, \delta \to 0$. A numerical expression for $\delta(s)$ is provided in Table 1. ${\hat \delta}(s)$ will be re-expressed in terms of other quantities below. 
\item[IV.] The  $\beta_\ell$ dependence in the coefficients is kept explicit (except for $J_{1s}$). This choice will be fundamental when approaching the massless limit to reduce the systematic associated to this limit.

\end{itemize}

\begin{figure}\begin{center}
\includegraphics[height=6cm,width=5cm]{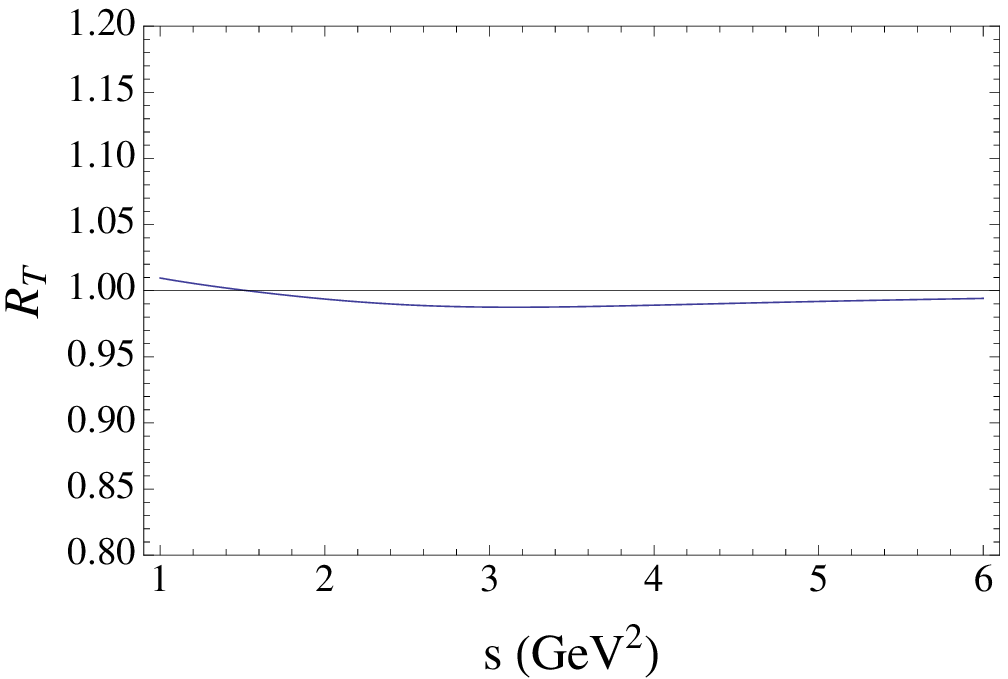}
\includegraphics[height=6cm,width=5cm]{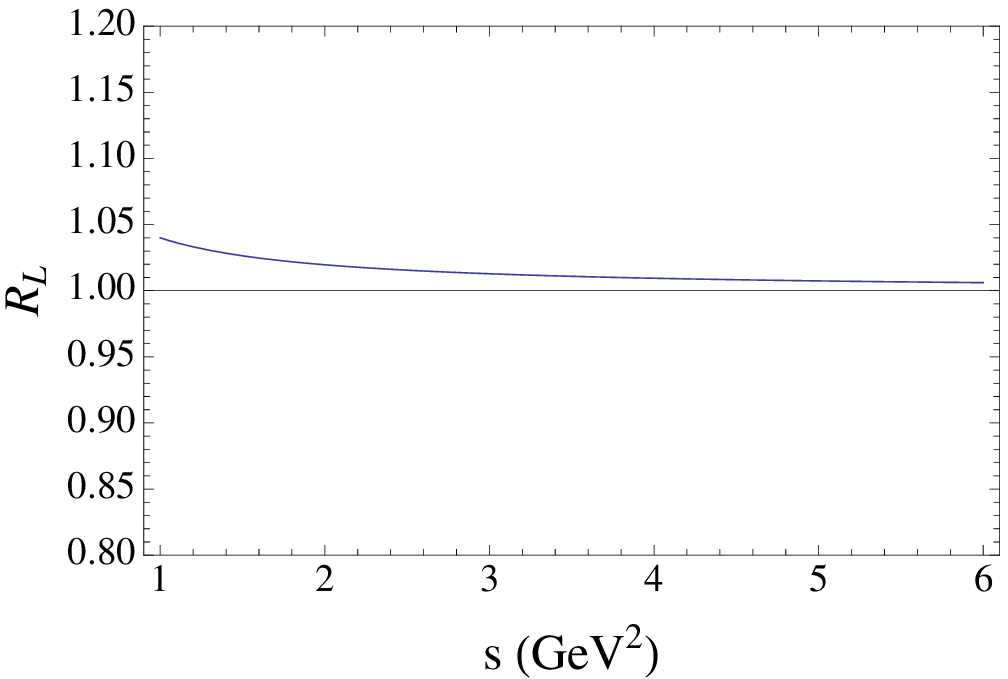}
\includegraphics[height=6cm,width=5cm]{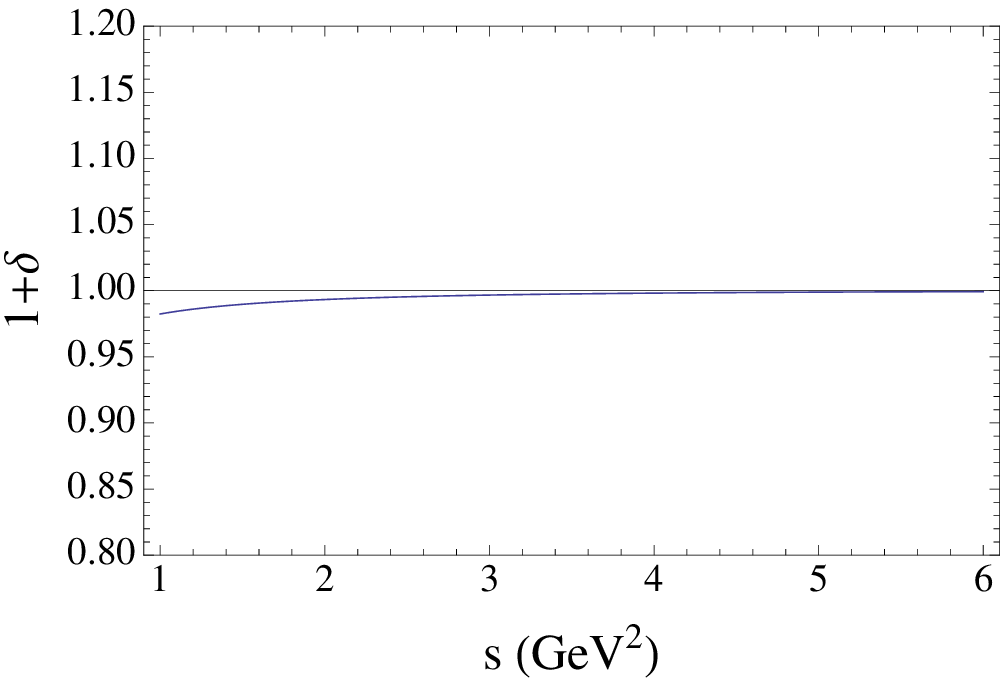}
\end{center}
\vspace{-0.4cm}
\caption{(left) ratio $R_T={\hat F}_T/{\tilde F}_T$  always below $1.2\%$ in SM, (middle) ratio $R_L={\hat F}_L/{\tilde F}_L$ with an error range  between $1\%$ to less than $4\%$ in the SM (right) $\delta$ represents the deviation from one of $\tilde F_L$+$\tilde F_T$. 
}
\label{SMplotsPs1}
\end{figure}

We will explore now in a bit more of detail the massless limit. It is clear that in the exact massive case one should strictly fit the data to four observables ${\hat F}_{L,T}$, ${\tilde F}_{L,T}$ apart from the six clean $P_i$. The ratios $R_T\equiv {\hat F}_T/{\tilde F}_T$  and $R_L\equiv {\hat F}_L/{\tilde F}_L$ are indeed  a direct measurement of the clean observables $M_1$ and $M_2$ defined in \cite{primary}:
$$ R_T=\frac{1}{3} \left(2 + (1+ 4 M_1)\beta_\ell^2 \right) \quad \quad R_L=1+ M_2 $$

The deviation from one of these ratios in the SM in the range $1$ to $6$ GeV$^2$ are shown in Fig.1. They are below $1.2 \%$ for $R_T$ and below $4 \%$ for $R_L$. Moreover, given the tight relation of these ratios with $M_{1,2}$ and the relatively small deviations found in these observables in the New Physics scenarios analyzed in \cite{primary}, a similar range  of deviations is expected also for New Physics. 

Finally, since it is very costly in statistics  to fit for the four observables ${\hat F}_{T,L}$, ${\tilde F}_{T,L}$ one could take the massless limit ($m_\ell \to 0$ everywhere) 
to reduce the number of observables to fit. 
However,  the exact massless limit could induce errors of order $4\%$ (around $q^2=1$ GeV$^2$) for instance if ${\hat F}_L={\tilde F}_L$ is used (see $R_L$ in Fig.1). For this reason we prefer to define an intermediate limit, that we call ``massless-improved limit". This consists 
in keeping the $\beta_\ell$ dependence of Eqs.(\ref{Jsintermsobs}) and reexpressing ${\hat F}_T$ and ${\hat F}_L$ in terms of ${\tilde F}_T$ in the following way
\bea \label{fthat}
{\hat F}_T &=& {\tilde F}_T + (R_T-1) {\tilde F}_T \equiv {\tilde F}_T + L_1 \\
{\hat F}_L &=& 1- {\tilde F}_T + { \delta} + (R_L-1)(1- {\tilde F}_T) + {\cal O}((R_L-1){ \delta}) \equiv 
1- {\tilde F}_T + L_2
\eea
where  ${\cal O}((R_L-1){ \delta})$ is much below the permille level and ${\hat \delta}(s)=L_1+L_2$. A numerical expression for $L_{1,2}$ in the SM can be found in Table 1. In this improved limit the contribution to ${\hat F}_{L,T}$ from lepton mass terms ($L_1, L_2$) in the SM is { always} below $1.3\% $ in all $q^2$-region from 1 to 6 GeV$^2$. Notice that the $P_i$ are not directly affected by this limit  since they are always defined by the same ${\tilde F}_{T,L}$, but they are only indirectly affected via the propagation of the error in the determination of ${\tilde F}_{T,L}$. The reason for the sizeable reduction from $4\%$ to near $1.3\%$ in the case of ${\hat F}_L$ is due to the partial cancellation between $ \delta$ and $(R_L-1) (1-{\tilde F}_T)$ in the SM (see Fig.1).
 This means that the error done by expressing the coefficients of the exact angular distribution Eqs.(\ref{Jsintermsobs}) and Eq.(\ref{redundant}) only in terms of ${\tilde F}_{T,L}$  (keeping the $\beta_\ell$ dependence but discarding the $L_{1,2}$ corrections) is { below} $1.3\%$. 



 Concerning the integrated observables there are  two important observations when including lepton masses. First notice  that 
$${\tilde F}_T \frac{d \Gamma_{K^*}}{dq^2}=\left({\tilde F}_T {\frac{d \Gamma_{K^*}}{dq^2}}\right)\Bigl|_{m_{\ell}=0} $$ and 
$$ \sqrt{{\tilde F}_T {\tilde F}_L} \frac{d \Gamma_{K^*}}{dq^2}=\left(\sqrt{{\tilde F}_T {\tilde F}_L} {\frac{d \Gamma_{K^*}}{dq^2}}\right)\Bigl|_{m_{\ell}=0} $$ 
 which means that these products are the same for the massive and massless case. Second important remark is that three of the $q^2$-dependent $P_i$ observables include a $\beta_\ell$ in their expression in terms of the $J_i$ coefficients and this can be problematic when defining integrated observables. The reason being that what is measured are the $J_i$ coefficients and not products like $\beta_\ell J_i$. Consequently, a small redefinition of three of these observables is required when including lepton masses in the integrated observables. This means that the integrated observables in the massive case naturally split in two categories: 
 \begin{itemize}
 \item First set: $P_1$, $P_3$ and $P_4^\prime$ their $q^2$-expression in terms of spin amplitudes and $J_i$ coefficients do not contain neither lepton mass terms nor $\beta_\ell$ prefactors. 
 \bea \label{pi} <P_1^{m_\ell\neq 0}>_{bin}&=& \frac{\int_{bin} J_3}{2 \int_{bin} J_{2s}}= \frac{\int_{bin} \beta_\ell^2 P_1 {\tilde F}_T \frac{d\Gamma_{K*}}{dq^2}}{\int_{bin} \beta_\ell^2 {\tilde F}_T \frac{d\Gamma_{K*}}{dq^2}}\nn \\ 
<P_3^{m_\ell\neq 0}>_{bin}&=& -\frac{\int_{bin} J_9}{4 \int_{bin} J_{2s}}= \frac{\int_{bin} \beta_\ell^2 P_3 {\tilde F}_T \frac{d\Gamma_{K*}}{dq^2}}{\int_{bin} \beta_\ell^2 {\tilde F}_T \frac{d\Gamma_{K*}}{dq^2}}\nn \\
<P_4^{\prime\, m_\ell\neq 0}>_{bin}&=&\frac{\int_{bin} J_4}{\sqrt{- \int_{bin} J_{2c} \int_{bin} J_{2s}}  }=\frac{\int_{bin} \beta_\ell^2 P_4^\prime\, \sqrt{{\tilde F}_T {\tilde F}_L} \frac{d\Gamma_{K*}}{dq^2}}{\sqrt{\int_{bin} \beta_\ell^2 {\tilde F}_T  \frac{d\Gamma_{K*}}{dq^2} \int_{bin} \beta_\ell^2  {\tilde F}_L \frac{d\Gamma_{K*}}{dq^2}}}.\eea
For these observables $P_1^{m_\ell \neq 0}=P_1$, $P_3^{m_\ell \neq 0}=P_3$ and $P_4^{\prime\, m_\ell \neq 0}=P_4^\prime$. Naturally, only when expressing them directly in terms of observables the $\beta_\ell$ prefactor arises. 

\item Second set: $P_2$, $P_5^\prime$, $P_6^\prime$ also their $q^2$-expression in terms of spin amplitudes  do not contain any lepton mass terms or $\beta_\ell$ prefactors. However their $q^2-$differential expression in terms of $J_i$ {\bf do} contain a  $\beta_\ell$ prefactor, which means that its corresponding massive definition  for the integrated observables requires a redefinition with a prefactor $1/\beta_\ell$:
%
\bea
<P_2^{m_\ell \neq 0}>_{bin}&=&\frac{\int_{bin} J_{6s}}{8 \int_{bin} J_{2s}  }=  \frac{\int_{bin} \beta_\ell^2  (P_2/\beta_\ell) {\tilde F}_T \frac{d\Gamma_{K*}}{dq^2}}{\int_{bin}\beta_\ell^2 {\tilde F}_T \frac{d\Gamma_{K*}}{dq^2}}\nn \\
<P_5^{\prime\, m_\ell \neq 0}>_{bin}&=&\frac{\int_{bin} J_5}{2 \sqrt{- \int_{bin} J_{2c} \int_{bin} J_{2s}}  }= \frac{\int_{bin} \beta_\ell^2 (P_5^\prime/\beta_\ell) \sqrt{{\tilde F}_T {\tilde F}_L} \frac{d\Gamma_{K*}}{dq^2}}
{\sqrt{\int_{bin} \beta_\ell^2 {\tilde F}_T  \frac{d\Gamma_{K*}}{dq^2} \int_{bin} \beta_\ell^2  {\tilde F}_L \frac{d\Gamma_{K*}}{dq^2}}}\nn\\
<P_6^{\prime \, m_\ell \neq 0}>_{bin}&=&-\frac{\int_{bin} J_7}{2 \sqrt{- \int_{bin} J_{2c} \int_{bin} J_{2s}}  }= \frac{\int_{bin}\beta_\ell^2 (P_6^\prime/\beta_\ell) \sqrt{{\tilde F}_T {\tilde F}_L} \frac{d\Gamma_{K*}}{dq^2}}
{\sqrt{\int_{bin} \beta_\ell^2 {\tilde F}_T  \frac{d\Gamma_{K*}}{dq^2} \int_{bin} \beta_\ell^2  {\tilde F}_L \frac{d\Gamma_{K*}}{dq^2}}}
\eea 
where $P_2^{m_\ell \neq 0}=P_2/\beta_\ell$, $P_5^{\prime\, m_\ell \neq 0}=P_5^\prime/\beta_\ell$ and $P_6^{\prime\, m_\ell \neq 0}=P_6^\prime/\beta_\ell$. 
 \end{itemize}
The last important point here is to compare the SM predictions for the massive integrated observables defined here $<P_i^{m_\ell \neq 0}>$ with the massless ones computed in \cite{DescotesGenon:2012zf}.  We will focus the comparison on the 1-bin  region between 1 to 6 GeV$^2$. For the first set 
due to the balance of $\beta_\ell^2$ factors between numerator and denominator of Eqs.(\ref{pi}) together with the slow variation of this factor with $s$, one should expect a tiny difference between the massive and the massless prediction. Indeed we found that the massive prediction in the SM is only $+0.8\%$ larger than the massless one for $P_1$ and $P_4^\prime$ and  below permille level for $P_3$. In the second set the comparison of the SM integrated observable in the range 1 to 6 GeV$^2$ in the massless  and massive case is $-0.6\%$, $+1.4\%$ and $+0.8\%$ for $P_2$, $P_5^\prime$ and $P_6^\prime$ respectively.
This implies that the massless predictions in the SM obtained in \cite{DescotesGenon:2012zf} are valid to an excellent approximation also in the massive case.



 
\begin{table}
\begin{center}
\small
\begin{tabular}{||c||c|c|c|c|c|c||}
\hline\hline
   & $1$ &   $s$ &  $s^2$ &  $s^3$
& $s^4$ & $s^5$  \\
\hline\hline
i &$0$ & $1$ & $2$ & $3$ & $4$ & $5$ \\
\hline\hline
{\rm dim} & GeV$^{0}$ &  GeV$^{-2}$ &  GeV$^{-4}$ &  GeV$^{-6}$ &  GeV$^{-8}$ &  GeV$^{-10}$ \\
 $\delta_{i}$ &
 -0.0498563 & +0.0502874 & -0.0228207 & +0.00540332 & -0.000644906 &
+0.000030558 \\
$R_{T\,i}$ & +1.02522 & -0.0074101 & -0.0142605 & +0.00717305 & -0.00120443 & +0.0000692328  \\
$ R_{L\,i}$ & +1.09905 & -0.09191 & +0.041595 & -0.00990168 & +0.00118789 & -0.0000565414 \\
$L_{1\,i}$ & +0.0201856 & -0.0273282 & +0.0135174 & -0.00331905 &
+0.000403496 & -0.0000193199 \\
$L_{2\,i}$ & -0.00598948 & +0.0255013 & -0.0149247 & +0.00389817 & -0.000488089 & +0.0000237576 \\
\hline\hline
\end{tabular}
\caption{Coefficients of the polynomial approximation to the deviation parameters $\delta$, $R_T$, $R_L$, $L_1$ and $L_2$, such that $\delta=\sum_{i=0,5} \delta_i s^i$, $R_T=\sum_{i=0,5} R_{T\,i} s^i$, $R_L=\sum_{i=0,5} R_{L\,i} s^i$, $L_1=\sum_{i=0,5} L_{1\, i} s^i$ and $L_2=\sum_{i=0,5} L_{2\, i} s^i$.}
\label{poly}
\end{center}
\end{table}

 In conclusion in this section we have proposed to use either the exact parametrization or an approximate parametrization so called "massless-improved" limit for the experimental fit (instead of the pure massless limit). This improved parametrization has the advantage of including the dominant contribution from lepton masses with an error when compared to the exact result that amounts only to a maximum of a  $1.3\%$  in the SM for the terms that contained ${\hat F}_{L,T}$ in the region between 1 to 6 GeV$^2$ while it is exact for the $q^2$ -dependent $P_i$ observables (up to the propagation of the error in the determination of ${\tilde F}_{T,L}$).  
 The practical consequence of this improved limit is to effectively reduce the basis of observables to one observable less  $\{\frac{d\Gamma_{K^*}}{dq^2},  {\tilde F}_T, {\tilde F}_L, P_{1,2,3},P_{4,5,6}^\prime \}$ (since we are taking $M_1 \to 0$). If one neglects in addition $\delta$ (keeping the $\beta_\ell$ dependence) by using ${\tilde F}_L\sim 1-{\tilde F}_T$ an additional error associated to this second approximation  (see Fig.1) should be added. In this case the previous basis gets reduced to eight observables. This step by step procedure will help to know precisely the size of the error associated to each approximation.
 Once this is done, next step is to compare with the theory predictions. We have shown here that the  comparison of the exact massive predictions in the SM with the massless predictions~\cite{DescotesGenon:2012zf} in the 1 to 6 GeV$^2$ bin for the integrated observables shows a deviation that in the worst case amounts to  $1.4\%$.


 

\section*{\normalsize II. Can one extract the observables $P_{1,2,3}$ and $P_{4,5,6}^\prime$ free from S-wave pollution?}

In order to answer this question two aspects have to be analyzed. On the one hand, the intrinsic character of the observable with respect to the S-wave pollution and, on the other, the extraction procedure of the observable.

Concerning the first aspect, one should distinguish the observables that does not suffer from S-wave pollution by construction from those that have an inherent structure that includes the pollution. 
It is clear  that 
the products
$${\tilde  F}_T \frac{d \Gamma_{K^*}}{dq^2},\quad {\tilde F}_L \frac{d \Gamma_{K^*}}{dq^2}, \quad P_{i} {\tilde F}_T \frac{d \Gamma_{K^*}}{dq^2}  \quad   P_{j}^\prime \sqrt{{\tilde F}_L {\tilde F}_T}  \frac{d \Gamma_{K^*}}{dq^2} 
$$
but also $P_i$, $P_j^\prime$ with $i=1,2,3$ and $j=4,5,6$ are all intrinsically free from this contamination.  Notice that these products enter in the definition  of  integrated observables.
 Consequently the corresponding integrated observables  are also intrinsically free from S-wave pollution.

On the contrary, for instance, 
 the $S_i$ observables \cite{th4} are normalized by the full differential decay distribution
 \eqa{ S_i = \frac{J_i+ {\bar J}_i}{\frac{d\Gamma_{full}}{dq^2} }}
  and they will be directly affected by the S-wave pollution  given that the full differential decay distribution is largely affected by this disease (see Eq.(\ref{full})). Any other observable normalized by $\frac{d\Gamma_{full}}{dq^2}$, like for example a longitudinal polarization fraction normalized by the full distribution (call it ${\bar F}_L$), will be strongly S-wave polluted, however the product ${\bar F}_L \frac{d\Gamma_{full}}{dq^2}$ is again free from pollution.

Concerning the extraction procedure, in the following
we will show that it is possible to extract these observables completely free from any S-wave pollution in the exact lepton mass case and, afterwards, we will present for completeness also an approximate solution using uniangular distributions.

\subsection*{\small  Exact solution including lepton mass corrections using folded distributions}

The crucial point here is to  determine the exact procedure to extract the $P_i$ observables. There are basically  two choices: uniangular distributions  or the full angular distribution. The major drawback of the former is that by integrating angles the distinct angular dependence of the unwanted ${\tilde J}_i$ is lost and mixing (and pollution) with the interesting $J_i$ is unavoidable. If, instead, the full angular distribution is used the major problem is that  there is not yet enough statistics to have access to the full distribution. However,  the clever idea presented in Ref.\cite{exp4} of using ``folded'' distributions, 
have the double advantage of increasing the statistics and focusing on a restricted set of angular coefficients. This allows to extract information from this mode in a more selective way than traditional uniangular distributions and, in particular,
 it can be used, as we will show, to isolate  the polluting ${\tilde J_i}$ terms. In this note we will exploit this trick to show that using  folded distributions one can measure all $P_i$ and $P_j^\prime$ free from any S-wave pollution thanks to their distinct angular dependence.



In Ref.~\cite{exp4}, the identification of $\phi \leftrightarrow \phi + \pi$ (when $\phi<0$) has been used to produce a ``folded'' angle $\hat \phi \in [0,\pi]$ in terms of which a (folded) differential decay rate $d\hat\Gamma = d\Gamma(\hat\phi)+d\Gamma(\hat\phi-\pi)$ is obtained\footnote{Other examples of singled folded distributions exhibiting dependencies on $P_{1,2,3}$ but with the angle $\theta_K$ instead can be found in Appendix B.}. This folding exploits the angular symmetries of the distribution and reduce to a subset  the coefficients  entering each folded distribution.
If we use the same folding procedure   including the extra terms coming from the pollution channel $B\to K_0^*(\to K \pi) l^+l^-$ one finds the well known structure for the angular distribution together with an extra piece with a distinguishable angular dependence
\eqa{\label{eqd1}
\frac{d^4\hat\Gamma}{dq^2\,d\!\cos\theta_K\,d\!\cos\theta_l\,d\hat\phi} &=&
\frac{9}{16\pi}
\bigg[ { J_{1c}} \cos^2{\theta_K} + { J_{1s}} (1 - \cos^2{\theta_K})
+ { J_{2c}} \cos^2{\theta_K} (2\cos^2{\theta_\ell}-1)\nn\\[2mm]
&&\hspace*{-1.5cm} + { J_{2s}} (1 - \cos^2{\theta_K}) (2\cos^2{\theta_\ell} - 1)
+ { J_3} (1 - \cos^2{\theta_K})(1 - \cos^2{\theta_\ell}) \cos{2\hat \phi}\nn\\[2mm]
&&\hspace*{-1.5cm} + { J_{6s}} (1 - \cos^2{\theta_K}) \cos{\theta_\ell}
+ { J_9} (1 - \cos^2{\theta_K})(1 - \cos^2{\theta_\ell}) \sin{2\hat \phi}
\bigg]X + W_1 }
 where
\eqa{&&W_1=\frac{1}{2\pi} \left[{\tilde J}_{1a}^c+{\tilde J}_{1b}^c \cos\theta_K+ \left({\tilde J}_{2a}^c+{\tilde J}_{2b}^c \cos \theta_K\right)(2\cos^2{\theta_\ell}-1)\right]}
The extra factor $W_1$   concentrates all the S-wave  pollution of this folded distribution. Its distinct angular dependence can be parametrized by
\bea \label{structureW} W_1=a_W+b_W (2 \cos^2 \theta_\ell -1) + c_W \cos\theta_K  + d_W \cos \theta_K (2 \cos^2 \theta_\ell -1)\eea
Fitting for these four parameters  ($a_W,b_W,c_W,d_W$)   one can measure and disentangle the S-wave pollution.\footnote{Notice that if lepton mass suppressed terms from the polluting decay $B\to K_0^*ll$ are disregarded $a_W=-b_W$ and $c_W=-d_W$} But, more importantly, the different angular dependence associated to the relevant coefficients involved in the definition of the $P_{1,2,3}$ permits an extraction of these observables in a way completely free from any S-wave pollution.  It is clear  from the definition of  $P_{1,2,3}$ in Eq.(\ref{props}) that the folded distribution provides access to $J_3$, $J_{6s}$ and $J_9$, numerators of $P_1$, $P_2$ and $P_3$ respectively. But also their common denominator $J_{2s}$ is identified with  the coefficient of $(1-\cos^2 \theta_K) (2 \cos^2 \theta_\ell -1)$. The normalization of ${\tilde A_T}^2$, $A_T^{({\rm re})}$ and $A_T^{({\rm im})}$ presented in \cite{damir} requires to measure in addition $J_{1c}$, unnecessary for the $P_{1,2,3}$.

This is more easily seen  writing down Eq.(\ref{eqd1}) in terms of observables using Eqs.(\ref{Jsintermsobs}). The following expression  generalizes Eq.(1) of the LHCb note \cite{exp4}  by including lepton mass corrections and the S-wave pollution in a compact form: 
\eqa{\label{eqd}
\frac{d^4\hat\Gamma}{dq^2\,d\!\cos\theta_K\,d\!\cos\theta_l\,d\hat\phi} &=&
\frac{9}{16\pi}
\bigg[ {\hat F_L} \cos^2{\theta_K} + \frac34{\hat F_T} (1 - \cos^2{\theta_K})
- \beta_\ell^2 {\tilde F_L} \cos^2{\theta_K} (2\cos^2{\theta_\ell}-1)\nn\\[2mm]
&&\hspace{-3.7cm} + \frac14 \beta_\ell^2 {\tilde F}_T (1 - \cos^2{\theta_K}) (2\cos^2{\theta_\ell} - 1)
+ \frac12 \beta_\ell^2 P_1{\tilde  F}_T (1 - \cos^2{\theta_K})(1 - \cos^2{\theta_\ell}) \cos{2\hat \phi}\nn\\[2mm]
&&\hspace{-3.7cm} + 2 \beta_\ell P_2{\tilde  F}_T (1 - \cos^2{\theta_K}) \cos{\theta_\ell}
- \beta_\ell^2 P_3 {\tilde F}_T (1 - \cos^2{\theta_K})(1 - \cos^2{\theta_\ell}) \sin{2\hat \phi}
\bigg]\, \frac{d\Gamma_{K*}}{dq^2} + W_1
}
or in the massless-improved limit:
\eqa{\label{eqd}
\frac{d^4\hat\Gamma}{dq^2\,d\!\cos\theta_K\,d\!\cos\theta_l\,d\hat\phi} &=&
\frac{9}{16\pi}
\bigg[(1- {\tilde F_T}) \cos^2{\theta_K} + \frac34{\tilde F_T} (1 - \cos^2{\theta_K})
- \beta_\ell^2 {\tilde F_L} \cos^2{\theta_K} (2\cos^2{\theta_\ell}-1)\nn\\[2mm]
&&\hspace{-3.2cm} + \frac14 \beta_\ell^2 {\tilde F}_T (1 - \cos^2{\theta_K}) (2\cos^2{\theta_\ell} - 1)
+ \frac12 \beta_\ell^2 P_1{\tilde  F}_T (1 - \cos^2{\theta_K})(1 - \cos^2{\theta_\ell}) \cos{2\hat \phi}\nn\\[2mm]
&&\hspace{-3.2cm} + 2 \beta_\ell P_2{\tilde  F}_T (1 - \cos^2{\theta_K}) \cos{\theta_\ell}
- \beta_\ell^2 P_3 {\tilde F}_T (1 - \cos^2{\theta_K})(1 - \cos^2{\theta_\ell}) \sin{2\hat \phi}
\bigg]\, \frac{d\Gamma_{K*}}{dq^2} + W_1\,
}
where the $\beta_\ell$ dependence is kept, ${\hat F}_L$ is traded by ${\tilde F}_T$ and the $L_{1,2}$ subleading terms have been dropped off. The error of this improved limit is below $1.3\%$ as discussed in the previous section.

There is also an alternative  that can be automatically applied to all the folded distributions discussed here and it is to use $\frac{d\Gamma_{full}}{dq^2}$ instead of $\frac{d\Gamma_{K*}}{dq^2}$, i.e, 
\eqa{\label{eqd}
\frac{d^4\hat\Gamma}{dq^2\,d\!\cos\theta_K\,d\!\cos\theta_l\,d\hat\phi} &=&
\frac{9}{16\pi}
\bigg[ {\bar{\hat F}_L} \cos^2{\theta_K} + \frac34{\bar{\hat F}_T} (1 - \cos^2{\theta_K})
- \beta_\ell^2 {\bar{\tilde F}_L} \cos^2{\theta_K} (2\cos^2{\theta_\ell}-1)\nn\\[2mm]
&&\hspace{-3.7cm} + \frac14 \beta_\ell^2 {\bar{\tilde F}_T} (1 - \cos^2{\theta_K}) (2\cos^2{\theta_\ell} - 1)
+ \frac12 \beta_\ell^2 P_1{\bar{\tilde  F}_T} (1 - \cos^2{\theta_K})(1 - \cos^2{\theta_\ell}) \cos{2\hat \phi}\nn\\[2mm]
&&\hspace{-3.7cm} + 2 \beta_\ell P_2{\bar{\tilde  F}_T} (1 - \cos^2{\theta_K}) \cos{\theta_\ell}
- \beta_\ell^2 P_3 {\bar{\tilde F}_T} (1 - \cos^2{\theta_K})(1 - \cos^2{\theta_\ell}) \sin{2\hat \phi}
\bigg]\, \frac{d\Gamma_{full}}{dq^2} + W_1
}
where all quantities with a bar are defined by $\bar Y= C Y$ with  $$C= \frac{\frac{d \Gamma_{K*}}{dq^2}}{\frac{d\Gamma_{full}}{dq^2}}$$ Notice that, as explained in the previous section, no S-wave pollution is introduced in this way in the relevant $q^2$ or integrated $P_i$ observables. Here the improved-massless limit  becomes defined by
\bea
{\bar{\hat F}_T}= C {\hat F}_T &=&   C {\tilde F}_T + (R_T-1) C {\tilde F}_T =  C {\tilde F}_T + C L_1 \eea
where the relative error induced by dropping  $C L_1$ in $C {{\tilde F}_T}$ is obviously the same as before, below $1.2\%$. However for 
${\bar{\hat F}_L}$ two possibilities are now open depending on the size of $C$. First possibility, if $|C-1|$ is comparable in size to $C L_2$ taking into account that $C-1$ is negative while $C L_2$ is positive a partial cancellation can take place and the best approach is to use 
\bea
{\bar{\hat F}_L}= C {\hat F}_L &=& 
1-  C{\tilde F}_T + C L_2 + (C-1)
\eea
In practice this require $|C-1|$ to be below $4\%$ to be a good choice (given that $L_2$ is below $1\%$). The second possibility is to take 
\bea
{\bar{\hat F}_L}= C {\hat F}_L &=& C{\tilde F}_L + (R_L-1) C {\tilde F}_L
\eea
In this case the relative error $R_L-1$ is independent of $C$ exactly as for ${\bar{\hat F}_R}$, however it rises up to $4\%$ at $q^2=1$ GeV$^2$ as can be seen in Fig. 1. This choice is safer because the error does not depend on $C$ but the high remaining error ($4\%$)  implies that the advantage of the massless-improved limit in front of the massless case  is lost.

At this point it is evident the power of this folding technique, in the massive case the full distribution would depend on 11 $J_i$ coefficients (only 10 of them are independent) plus 8 ${\tilde J}_i$ coefficients from the companion decay. Instead, the folded distribution depends only on seven $J_i$ coefficients (or seven observables ${\hat F}_T,{\hat F}_L,{\tilde F}_T,{\tilde F}_L, P_1, P_2, P_3$  ) and four ${\tilde J}_i$ ($a_W,b_W,c_W,d_W$). If the improved-massless limit is used the number of observables reduces  to six: $d\Gamma_{K^*}/{dq^2},{\tilde F}_T$, ${\tilde F}_L$, $P_1$, $P_2$, $P_3$ plus the four from ${\tilde J}_i$. 


Notice also that the prefactor $X=\int dm_{K\pi}^2 |BW_{K^*}(m_{K\pi}^2)|^2$ cancels out exactly in the  $P_i$ given their definition in terms of ratios of coefficients. 


One can also exploit further this technique and find  several doubled-folded distribution showing different dependencies on the $P_{1,2,3}$ observables and exhibiting common/different S-wave pollutions. For instance, two possible foldings of  
the angles $\phi$ and $\theta_\ell$ are 
\begin{itemize}

\item[IV.] Identifying $\phi\leftrightarrow \phi+\pi$ when $\phi<0$ and $\theta_\ell \leftrightarrow \theta_\ell- \frac{\pi}{2}$ when $\theta_\ell>\frac{\pi}{2}$ with $\hat \phi \in [0,\pi]$ and $\hat \theta_\ell \in [0,\pi/2]$ such that the folded differential distribution is 
$$d\hat\Gamma=d\Gamma(\hat\phi,\hat\theta_\ell,\theta_K)+d\Gamma(\hat\phi,\hat\theta_\ell+\frac{\pi}{2},\theta_K)+
d\Gamma(\hat\phi-\pi,\hat\theta_\ell,\theta_K)+d\Gamma(\hat\phi-\pi,\hat\theta_\ell+\frac{\pi}{2},\theta_K)$$
corresponding to 
\bea \frac{d^4\hat\Gamma}{dq^2\,d\!\cos\theta_K\,d\!\cos\hat\theta_l\,d\hat\phi}&=&\frac{9}{32\pi} \left[4 {\hat F}_L \cos^2\theta_K+3 {\hat F}_T \sin^2\theta_K +  {\tilde F}_T \sin^2\theta_K (\beta_\ell^2 P_1 \cos 2\hat\phi   
 \right.\nn\\ &&\left.
 + 4 \beta_\ell P_2 (\cos\hat\theta_\ell-\sin\hat\theta_\ell)
  - 2 \beta_\ell^2 P_3 \sin 2\hat\phi) \right] \frac{d\Gamma_{K*}}{dq^2} +W_4
\eea
where $$W_4=\frac{1}{\pi} \left[{\tilde J}_{1a}^c+{\tilde J}_{1b}^c \cos \theta_K \right]$$

\item[V.] Identifying $\phi\leftrightarrow \phi+\pi$ when $\phi<0$ and $\theta_\ell \leftrightarrow \pi -\theta_\ell $  when $\theta_\ell > \frac{\pi}{2}$ with $\hat \phi \in [0,\pi]$ and $\hat \theta_\ell \in [0,\pi/2]$ and the folded distribution
$$d\hat\Gamma=d\Gamma(\hat\phi,\hat\theta_\ell,\theta_K)+d\Gamma(\hat\phi,\pi-\hat\theta_\ell,\theta_K)+
d\Gamma(\hat\phi-\pi,\hat\theta_\ell,\theta_K)+d\Gamma(\hat\phi-\pi,\pi-\hat\theta_\ell,\theta_K)$$
corresponding to
\bea \frac{d^4\hat\Gamma}{dq^2\,d\!\cos\theta_K\,d\!\cos\hat\theta_l\,d\hat\phi}&=&
\frac{9}{32\pi} \left[4 \cos^2\theta_K ({\hat F}_L- \beta_\ell^2 {\tilde F}_L \cos 2\hat\theta_\ell) + \sin^2 \theta_K (3 {\hat F}_T+
\beta_\ell^2 {\tilde F}_T \cos 2\hat\theta_\ell)+\right.\nn \\&&
\left. +2 \beta_\ell^2 \sin^2\theta_K {\tilde F}_T (P_1 \cos 2\hat\phi-2 P_3 \sin 2\hat\phi) \sin^2\hat\theta_\ell)\right] \frac{d\Gamma_{K*}}{dq^2}  + W_5\eea
where $W_5=2 W_1(\hat\phi,\hat\theta_\ell,\theta_K)$. 
\end{itemize}
Other double foldings with  the angles $\phi$ and  $\theta_K$ that offer a different dependence on the 
observables $P_{1,2,3}$ can be found in Appendix B.

Finally, we show how the pollution can be also disentangled from the  $P_{4,5,6}^\prime$ observables using this folding technique. Here we present different possibilities:
\begin{itemize}
\item [X.] Identifying the angle $\phi \leftrightarrow -\phi$ when $\phi<0$ with $\hat \phi \in [0,\pi]$ and the differential folded distribution is now $d\hat\Gamma=d\Gamma(\hat\phi,\theta_\ell,\theta_K)+d\Gamma(-\hat\phi,\theta_\ell,\theta_K)$ where
\bea 
\frac{d^4\hat\Gamma}{dq^2\,d\!\cos\theta_K\,d\!\cos\theta_l\,d\hat\phi}&=&
\frac{9}{64\pi}\left[ 4 \cos^2\theta_K ({\hat F}_L -\beta_\ell^2 {\tilde F}_L \cos 2 \theta_\ell)+(3{\hat F}_T + \beta_\ell^2 {\tilde F}_T \cos 2 \theta_\ell) \sin^2\theta_K+ \right. \nonumber \\
&&\left.\hspace*{-1.7cm}+
2 {\tilde F}_T (\beta_\ell^2 P_1 \cos 2\hat\phi \sin^2 \theta_K \sin^2\theta_\ell + 4 \beta_\ell P_2 \cos\theta_\ell \sin^2\theta_K)+\right.
\nn \\ &&\left.\hspace*{-1.7cm}
 +
2 \sqrt{{\tilde F}_L {\tilde F}_T} ( \beta_\ell^2 P_4^\prime \sin 2 \theta_K \sin 2 \theta_\ell + 2 \beta_\ell P_5^\prime \sin 2 \theta_K \sin \theta_\ell) \cos \hat\phi \right]\frac{d\Gamma_{K*}}{dq^2}  + W_{10}\quad  \quad
\eea
with $$W_{10}=\frac{1}{2\pi}\left[{\tilde J}_{1a}^c+{\tilde J}_{2a}^c \cos 2\theta_\ell + 
\cos \theta_K ({\tilde J}_{1b}^c + {\tilde J}_{2b}^c \cos 2 \theta_\ell)+
\cos\hat\phi ({\tilde J}_5+ 2 {\tilde J}_4\cos\theta_\ell)\sin \theta_K \sin\theta_\ell
\right]
$$
\item[XI.] Identifying the angle $\theta_\ell\leftrightarrow \pi-\theta_\ell$ when $\theta_\ell>\frac{\pi}{2}$ with $\hat \theta_\ell \in [0,\pi/2]$ and the folded distribution $d\hat\Gamma=d\Gamma(\phi,\hat\theta_\ell,\theta_K)+d\Gamma(\phi,\pi-\hat\theta_\ell,\theta_K)$ is then 
\bea \frac{d^4\hat\Gamma}{dq^2\,d\!\cos\theta_K\,d\!\cos\hat\theta_l\,d\phi}&=&\frac{9}{64\pi}\left[ 4 \cos^2\theta_K ({\hat F}_L -\beta_\ell^2 {\tilde F}_L \cos 2 \hat\theta_\ell)+ 
 (3{\hat F}_T + \beta_\ell^2 {\tilde F}_T \cos 2 \hat\theta_\ell) \sin^2\theta_K \right. \nn \\ &&+ 2\beta_\ell^2 {\tilde F}_T ( P_1 \cos 2\phi \sin^2 \theta_K \sin^2\hat\theta_\ell - 2 P_3 \sin 2 \phi \sin^2\hat\theta_\ell \sin^2\theta_K)+
\nn \\ &&\left.
 +
4\beta_\ell \sqrt{{\tilde F}_L {\tilde F}_T} (  P_5^\prime \cos\phi  - P_6^\prime \sin\phi)\sin 2 \theta_K \sin \hat\theta_\ell  \right]\frac{d\Gamma_{K*}}{dq^2}  + W_{11}\nn
\eea
where $$W_{11}=\frac{1}{2\pi}\left[{\tilde J}_{1a}^c+{\tilde J}_{2a}^c \cos 2\hat\theta_\ell + 
\cos \theta_K ({\tilde J}_{1b}^c + {\tilde J}_{2b}^c \cos 2 \hat\theta_\ell)+
 ({\tilde J}_5 \cos\phi+  {\tilde J}_7 \sin\phi)\sin \theta_K \sin\hat\theta_\ell
\right]
$$

\begin{table}
\begin{center}
\small
\begin{tabular}{||c|c|c|c|c|c||}
\hline\hline
$P_1$ &   $P_2$ &  $P_3$ &  $P_4^\prime$
& $P_5^\prime$ & $P_6^\prime$  \\
\hline
(1)-(13) & (1)-(4), (6)-(10),(12),(13) & (1)-(7),(11)-(12) & (10) & (10)-(13) &
(11),(12) \\
\hline\hline
\end{tabular}
\caption{Sensitivities to each $P_i$ observable from the set of 13 folding distributions discussed.}
\label{WCSM}
\end{center}
\end{table}

\item[XII.] Identifying $\theta_\ell \leftrightarrow \theta_\ell-\frac{\pi}{2}$ when $\theta_\ell>\frac{\pi}{2}$ with $\hat \theta_\ell \in [0,\pi/2]$ and the folded distribution $d\hat\Gamma=d\Gamma(\phi,\hat\theta_\ell,\theta_K)+d\Gamma(\phi,\hat\theta_\ell+\frac{\pi}{2},\theta_K)$ is
\bea 
\frac{d^4\hat\Gamma}{dq^2\,d\!\cos\theta_K\,d\!\cos\hat\theta_l\,d\phi}&=&
\frac{9}{32\pi}\left[ 
\frac{1}{4} ( 4 {\hat F}_L + 3 {\hat F}_T+ (4 {\hat F}_L - 3 {\hat F}_T)\cos 2 \theta_K)+
 \right. \nn \\ 
&&\hspace*{-3cm}+  {\tilde F}_T (\frac{1}{2} \beta_\ell^2 P_1 \cos 2\phi \sin^2 \theta_K + 2\beta_\ell P_2 \sin^2\theta_K (\cos\hat\theta_\ell-\sin\hat\theta_\ell)- \beta_\ell^2 P_3 \sin 2 \phi \sin^2\theta_K)
\nn \\ &&\left.
\hspace*{-3cm} +
\beta_\ell \sqrt{{\tilde F}_L {\tilde F}_T} (  P_5^\prime \cos\phi  - P_6^\prime \sin\phi)\sin 2 \theta_K (\sin \hat\theta_\ell
 +\cos\hat\theta_\ell)  \right]\frac{d\Gamma_{K*}}{dq^2}  + W_{12}\nn
\eea
where $$W_{12}=\frac{1}{4\pi}\left[ 2{\tilde J}_{1a}^c+2{\tilde J}_{1b}^c \cos \theta_K + 
 ({\tilde J}_5 \cos\phi+  {\tilde J}_7 \sin\phi)\sin \theta_K (\cos\hat\theta_\ell+\sin\hat\theta_\ell)
\right]
$$
\end{itemize}
One last example of two-folded angle distribution sensitive to $P_5^\prime$ comes from identifying $\phi \leftrightarrow -\phi$ when $\phi<0$ and $\theta_\ell \leftrightarrow \theta_\ell - \frac{\pi}{2}$ when $\theta_\ell>\frac{\pi}{2}$. The corresponding combination is 
$$d\hat\Gamma=d\Gamma(\hat\phi,\hat\theta_\ell,\theta_K)+d\Gamma(\hat\phi,\hat\theta_\ell+\frac{\pi}{2},\theta_K)+
d\Gamma(-\hat\phi,\hat\theta_\ell,\theta_K)+d\Gamma(-\hat\phi,\hat\theta_\ell+\frac{\pi}{2},\theta_K)$$
where 
\bea \frac{d^4\hat\Gamma}{dq^2\,d\!\cos\theta_K\,d\!\cos\hat\theta_l\,d\hat\phi}&=&
\frac{9}{32\pi}\left[ 
\frac{1}{2} ( 4 {\hat F}_L + 3 {\hat F}_T+ (4 {\hat F}_L - 3 {\hat F}_T)\cos 2 \theta_K)
+ \right. \nn \\ &&\left.
+ 2 \beta_\ell \sqrt{{\tilde F}_L {\tilde F}_T}  P_5^\prime \cos\hat\phi \sin 2 \theta_K (\sin \hat\theta_\ell
 +\cos\hat\theta_\ell) + \right.
\nn \\ &&\left. +
 {\tilde F}_T \sin^2 \theta_K (\beta_\ell^2 P_1 \cos 2\hat\phi  + 4 \beta_\ell P_2  (\cos\hat\theta_\ell-\sin\hat\theta_\ell))
  \right]\frac{d\Gamma_{K*}}{dq^2} + W_{13}\nn
\eea
with $$W_{13}=\frac{1}{2\pi}\left[ 2 {\tilde J}_{1a}^c+2{\tilde J}_{1b}^c \cos\theta_K + {\tilde J}_5 \cos\hat\phi \sin\theta_K (\cos\hat\theta_\ell+\sin\hat\theta_\ell)\right]$$


\bigskip

Of course, using the same procedure also the integrated observables   
can be extracted cleanly from the folded distributions
in a way completely free from any S-wave pollution.

\subsection*{\small Approximate solution using uniangular distributions}

In Ref\cite{damir}  the 'unusual' normalization factor $3 J_{1s}-J_{2s}$ of the transverse asymmetries is obtained from the uniangular distribution  
\bea\label{eq:bk}
{d^2\Gamma\over dq^2 d\cos\theta_K} = a_{\theta_K} (q^2)+ b_{\theta_K} (q^2)\cos\theta_K + c_{\theta_K} (q^2)\cos^2\theta_K\,, 
\eea
where the coefficient functions  are easily obtained integrating the angular distribution Eq.(\ref{dist}) over the angles $\phi$ and $\theta_\ell$. In particular, in our notation  one finds
\begin{align}\label{delta}
a_{\theta_K}(q^2) &= {\tilde J}_{1a}^c - \frac{1}{3} {\tilde J}_{2a}^c +\frac{3}{8} \left(3 J_{1s}-J_{2s} \right) X
= {\tilde J}_{1a}^c - \frac{1}{3} {\tilde J}_{2a}^c +  \frac{3}{32} \left(9 {\hat F}_T - \beta^2 {\tilde F}_T \right) \frac{d\Gamma_{K^*}}{dq^2} \nonumber \\ &\equiv \frac{3}{32} \left(9 {\hat F}_T - \beta^2 {\tilde F}_T \right) \left(1+ \Delta(q^2) \right) \frac{d\Gamma_{K^*}}{dq^2} 
\end{align}which corresponds exactly to Eq.(26) of Ref.\cite{damir} and defines in our notation the function $\Delta(q^2)$ introduced in Ref.\cite{damir}.
This function $\Delta(q^2)$ is the responsible for the deviation up to $23 \%$ around $q^2=2$ GeV$^2$.
Indeed it is not surprising this huge effect.  It is immediate to see from the definition of $\Delta(q^2)$ in \cite{damir} in terms of transversity amplitudes that it is 
proportional to $|{\cal M}_{0}^{\prime L}|^2+|{\cal M}_{0}^{\prime R}|^2$ (see Ref.\cite{damir} for definitions) which is the source of this pollution and large deviation. Notice that with the procedure described in \cite{damir}  this pollution is not suppressed by any $m_\ell^2/q^2$ prefactor.

However, if one uses instead the observables $P_1,P_2$ and $P_3$ defined as in Eq.(\ref{props}) one can design a procedure to extract their normalization $J_{2s}$ using uniangular distributions  with a lepton mass suppressed S-wave pollution. In the region between 1 to 6 GeV$^2$ this amounts to a much small pollution than the one found using $\Delta(q^2)$ coming from unsuppressed lepton mass terms.



We should emphasize that using uniangular distributions is not the best choice, since as we have shown  folded distributions allow for an exact extraction of the $P_i$ free from any pollution. Still if one insist in using uniangular distributions a procedure to obtain this normalization is easily constructed using the distributions on $\theta_\ell$ and $\phi$.
The idea is to combine  the two uniangular distributions
\bea
{d^2\Gamma\over dq^2 d \cos\theta_\ell}&=&a_{\theta_\ell} + b_{\theta_\ell} \cos \theta_\ell + c_{\theta_\ell} \cos^2 \theta_\ell \\
{d^2\Gamma\over dq^2 d \phi}&=&a_\phi  + b_\phi \cos \phi + c_\phi \sin \phi + d_\phi \cos 2 \phi + e_\phi \sin 2 \phi \eea
where  the coefficients can be trivially obtained by integrating the corresponding angles. Two possible combinations relevant for our purposes are:
\bea
-a_{\theta_\ell}+ 3 \pi \frac{1+\beta_\ell^2}{3 + \beta_\ell^2} a_\phi&=&K_1 \nn \\
a_{\theta_\ell}+ \frac{1 + \beta_\ell^2}{2 \beta_\ell^2} c_{\theta_\ell}&=&K_2
\eea
where 
\bea K_1&=&\frac{1}{8 (3+\beta_\ell^2)} \left[ 3 \beta_\ell^2 \left( 3 {\hat F}_T+ {\tilde F}_T+ 2 ({\hat F}_L -{\tilde F}_L) \right) \frac{d \Gamma_{K^*}}{dq^2} + 16 (\beta_\ell^2 {\tilde J}_{1a}^c+{\tilde J}_{2a}^c) \right] \nn \\
K_2 &=& \frac{3 + \beta_\ell^2}{2 \beta_\ell^2} K_1
\eea
In turn they can be written as
$$K_1= \frac{3}{2} X J_{2s} + {\cal O}\left(\frac{m_\ell^2}{q^2}\right) \quad K_2=3 X J_{2s} + {\cal O}\left(\frac{m_\ell^2}{q^2}\right)
$$
which shows that both combinations allow the extraction of the normalization coefficient $J_{2s}$  with an error given by terms of order ${\cal O}(m_\ell^2/q^2)$. 

%
The influence of the  $ {\cal O}(m_l^2/q^2)$ term    is moderate in the region of interest between 1 and 6 GeV$^2$ (below 1 GeV$^2$ more difficult problems and uncertainties arises).



\subsection*{Conclusions}
We have shown  that the $P_i$ observables are not only intrinsically clean observables from S-wave pollution point of view, but also that they can be extracted  using folded distributions
in a way completely free from any S-wave contamination coming from the companion decay $B\to K_0^* l^+l^-$, including all lepton mass corrections. The same conclusion applies to the corresponding  integrated observables. 

We have also defined a massless-improved limit for the differential distribution, an intermediate stage between the massive and the massless case, that effectively reduces the number of observables at a relatively low cost: an error below $1.3\%$ in the SM in the region between 1 to 6 GeV$^2$. Also a comparison of the SM prediction for the integrated observables $<P_i>_{q^2=1-6 {\rm GeV}^2} $ in the massive and massless case shows a discrepancy below  
$1.4\%$ in the SM, implying that the SM massless predictions for the integrated observables given in\cite{DescotesGenon:2012zf}  are an excellent approximation.

Both results are important  experimentally to get a better control or even remove the systematics associated to S-wave pollution and lepton masses.


\bigskip
\bigskip
\noindent{\bf Acknowledgements}\\
I acknowledge Nicola Serra for enlighting discussions on the folding technique, Thomas Blake and Damir Becirevic for useful e-mail exchange. I also acknowledge  financial support from FPA2011-25948, SGR2009-00894 and Julia, Jana and Muntsa for being so patient with me during August.
\bigskip

\appendix
\subsection*{Appendix A. ${\tilde J}_i$ definitions}
The definition of the ${\tilde J}_i$  derive directly from the full angular distribution given in \cite{damir}
\bea
{\tilde J}_{1a}^{c}&=&{\cal I}_{1}^{c\prime}(q^2)\int |BW_{K_0^*}(m_{K\pi}^2)|^2 dm_{K\pi}^2 \nn \\
{\tilde J}_{2a}^{c}&=&{\cal I}_{2}^{c\prime}(q^2)\int |BW_{K_0^*}(m_{K\pi}^2)|^2 dm_{K\pi}^2 \nn \\
{\tilde J}_{1b}^{c}&=&2 \sqrt{3} \int  {\rm Re} \left[{\cal I}_{1}^{c\prime\prime} BW_{K_0^*}(m_{K\pi}^2) BW_{K*}^\dagger (m_{K\pi}^2)\right]dm_{K\pi}^2 \nn \\
{\tilde J}_{2b}^{c}&=&2 \sqrt{3} \int  {\rm Re} \left[{\cal I}_{2}^{c\prime\prime}(q^2) BW_{K_0^*}(m_{K\pi}^2) BW_{K^*}^\dagger (m_{K\pi}^2)\right]dm_{K\pi}^2 \nn \\
{\tilde J}_{4}&=&2\sqrt{3} \int  {\rm Re} \left[{\cal I}_{4}^{\prime\prime}(q^2) BW_{K_0^*}(m_{K\pi}^2) BW_{K^*}^\dagger (m_{K\pi}^2)\right]dm_{K\pi}^2 \nn \\
{\tilde J}_{5}&=&2\sqrt{3} \int  {\rm Re} \left[{\cal I}_{5}^{\prime\prime}(q^2) BW_{K_0^*}(m_{K\pi}^2) BW_{K*}^\dagger (m_{K\pi}^2)\right]dm_{K\pi}^2 \nn \\
{\tilde J}_{7}&=&2\sqrt{3} \int  {\rm Im} \left[{\cal I}_{7}^{\prime\prime}(q^2) BW_{K_0^*}(m_{K\pi}^2) BW_{K*}^\dagger (m_{K\pi}^2)\right]dm_{K\pi}^2 \nn \\
{\tilde J}_{8}&=&2\sqrt{3} \int  {\rm Im} \left[{\cal I}_{8}^{\prime\prime}(q^2) BW_{K_0^*}(m_{K\pi}^2) BW_{K*}^\dagger (m_{K\pi}^2)\right]dm_{K\pi}^2 
\eea
We refer the reader to \cite{damir} for definitions of ${\cal I}_i$ and Breight-Wigner resonances $BW_{K_0^*}(m_{K\pi}^2)$, $BW_{K^*} (m_{K\pi}^2)$. See also \cite{lu}.
\bigskip

\subsection*{Appendix B. More folded distributions}

\subsubsection*{Observables $P_{1,2,3}$}

$\bullet$ Single folded distributions:The angle $\theta_K$ can be folded in two different ways.
\begin{itemize}
\item[II.] The first comes from the identification $\theta_K \leftrightarrow \pi-\theta_K$ when $\theta_K>\frac{\pi}{2}$ with $\hat \theta_K \in [0,\pi/2]$ implying the combination $d\hat\Gamma=d\Gamma(\phi,\theta_\ell,\hat\theta_K)+d\Gamma(\phi,\theta_\ell,\pi-\hat\theta_K)$ corresponding to
\bea \frac{d^4\hat\Gamma}{dq^2\,d\!\cos\hat\theta_K\,d\!\cos\theta_l\,d\phi}&=&
\frac{9}{64\pi}
\left[4 \cos^2 \hat\theta_K ({\hat F}_L- \beta_\ell^2{\tilde F}_L \cos 2 \theta_\ell)
+\sin^2 \hat\theta_K (3 {\hat F}_T + \beta_\ell^2 {\tilde F}_T \cos 2 \theta_\ell)+ \right. \nn\\ &&
+ 2 {\tilde F}_T ( \beta_\ell^2 P_1 \cos 2\phi\sin^2\hat\theta_K\sin^2\theta_\ell + 4 \beta_\ell P_2 \cos\theta_\ell \sin^2 \hat\theta_K  \nn \\
&&\left. -2\beta_\ell^2 P_3 \sin 2 \phi \sin^2 \hat\theta_K \sin^2 \theta_\ell) \right]\frac{d\Gamma_{K*}}{dq^2}  + W_2\eea
where $$W_2=\frac{1}{2\pi}\left[{\tilde J}_{1a}^c+{\tilde J}_{2a}^c \cos 2\theta_\ell + (\cos\phi ({\tilde J}_5+ 2 {\tilde J}_4\cos\theta_\ell)+({\tilde J}_7+ 2 {\tilde J}_8 \cos\theta_\ell)\sin\phi)\sin\hat\theta_K\sin\theta_\ell\right]
$$
\item[III.] The second is $\theta_K \leftrightarrow \theta_K- \frac{\pi}{2}$ when $\theta_K>\frac{\pi}{2}$ with $\hat \theta_K \in [0,\pi/2]$ and the combination is $d\hat\Gamma=d\Gamma(\phi,\theta_\ell,\theta_K)+d\Gamma(\phi,\theta_\ell,\theta_K+\frac{\pi}{2})$ given by
\bea  &&\frac{d^4\hat\Gamma}{dq^2\,d\!\cos\hat\theta_K\,d\!\cos\theta_l\,d\phi}=\frac{9}{128\pi} \left[4 {\hat F}_L + 3 {\hat F}_T + (-4 {\tilde F}_L+{\tilde F}_T)\beta_\ell^2\cos 2 \theta_\ell+ \right. \nn\\
&&\left. +2 {\tilde F_T} (\beta_\ell^2 P_1 \cos2\phi\sin^2\theta_\ell+4 \beta_\ell P_2 \cos\theta_\ell - 2 \beta_\ell^2 P_3 \sin 2\phi \sin^2\theta_\ell)
\right] \frac{d\Gamma_{K*}}{dq^2}  + W_3\eea
where \bea W_3&=&\frac{1}{4\pi}\left[2 ({\tilde J}_{1a}^c+{\tilde J}_{2a}^c \cos2\theta_\ell)+({\tilde J}_{1b}^c+{\tilde J}_{2b}^c\cos 2 \theta_\ell)(\cos\hat\theta_K-\sin\hat\theta_K)+(\cos\phi ({\tilde J}_5+ 2 {\tilde J}_4 \cos\theta_\ell) \right. \nn\\
 &&\left. + ({\tilde J}_7 + 2 {\tilde J}_8 \cos\theta_\ell)\sin\phi )(\cos \hat\theta_K + \sin\hat\theta_K)\sin\theta_\ell\right]
\eea
\end{itemize}

$\bullet$ Double folded distributions:
\medskip

VI. Identifying $\phi\leftrightarrow \phi+\pi$ when $\phi<0$ and $\theta_K \leftrightarrow \theta_K- \frac{\pi}{2}$ when $\theta_K>\frac{\pi}{2}$ with $\hat \phi \in [0,\pi]$ and $\hat \theta_K \in [0,\pi/2]$ with a combined distribution
$$d\hat\Gamma=d\Gamma(\hat\phi,\theta_\ell,\hat\theta_K)+d\Gamma(\hat\phi,\theta_\ell,\hat\theta_K+\frac{\pi}{2})+
d\Gamma(\hat\phi-\pi,\theta_\ell,\hat\theta_K)+d\Gamma(\hat\phi-\pi,\theta_\ell,\hat\theta_K+\frac{\pi}{2})$$ given by
\bea \frac{d^4\hat\Gamma}{dq^2\,d\!\cos\hat\theta_K\,d\!\cos\theta_l\,d\hat\phi}&=&
\frac{9}{64\pi} \left[ 4 {\hat F}_L+3 {\hat F}_T + 
(- 4 {\tilde F}_L+{\tilde F}_T)\beta_\ell^2  \cos 2\theta_\ell + 8 {\tilde F}_T\beta_\ell P_2 \cos \theta_\ell+ \right. \nn\\
&&\left.+ 2 \beta_\ell^2 {\tilde F}_T (P_1 \cos 2\hat\phi
- 2  P_3 \sin 2\hat \phi) \sin^2\theta_\ell \right] \frac{d\Gamma_{K*}}{dq^2}  + W_6 
\eea
where 
$$W_6=\frac{1}{2\pi} \left[2({\tilde J}_{1a}^c+{\tilde J}_{2a}^c \cos2\theta_\ell)+({\tilde J}_{1b}^c+{\tilde J}_{2b}^c \cos 2 \theta_\ell) (\cos\hat \theta_K-\sin\hat \theta_K)\right]$$

VII. Identifying $\phi\leftrightarrow \phi+\pi$ when $\phi<0$ and $\theta_K \leftrightarrow \pi -\theta_K $ when $\theta_K>\frac{\pi}{2}$ with $\hat \phi \in [0,\pi]$ and $\hat \theta_K \in [0,\pi/2]$ and the combined distribution 
$$d\hat\Gamma=d\Gamma(\hat\phi,\theta_\ell,\hat\theta_K)+d\Gamma(\hat\phi,\theta_\ell,\pi-\hat\theta_K)+
d\Gamma(\hat\phi-\pi,\theta_\ell,\hat\theta_K)+d\Gamma(\hat\phi-\pi,\theta_\ell,\pi-\hat\theta_K)$$
where 
\bea 
\frac{d^4\hat\Gamma}{dq^2\,d\!\cos\hat\theta_K\,d\!\cos\theta_l\,d\hat\phi}=
 \frac{9}{32\pi} \left[ 4 \cos^2\hat\theta_K ({\hat F}_L- \beta_\ell^2 {\tilde F}_L \cos 2\theta_\ell) + \sin^2 \hat\theta_K (3 {\hat F}_T + \beta_\ell^2 {\tilde F}_T \cos 2 \theta_\ell)+\right.&&\nn\\ \left.
+2   \sin^2 \hat\theta_K {\tilde F}_T \left(\beta_\ell^2 (P_1 \cos 2\hat\phi  - 2 P_3 \sin 2 \hat\phi) \sin^2 \theta_\ell + 4 \beta_\ell P_2 \cos \theta_\ell\right) \right] \frac{d\Gamma_{K*}}{dq^2}  + W_7&&\quad\quad
\eea
with $$W_7=\frac{1}{\pi} \left[{\tilde J}_{1a}^c+{\tilde J}_{2a}^c \cos 2\theta_\ell \right]$$

VIII. Identifying $\phi\leftrightarrow -\phi$ when $\phi<0$ and $\theta_K \leftrightarrow \pi -\theta_K $ when $\theta_K>\frac{\pi}{2}$ with $\hat \phi \in [0,\pi]$ and $\hat \theta_K \in [0,\pi/2]$ with the combined distribution 
$$d\hat\Gamma=d\Gamma(\hat\phi,\theta_\ell,\hat\theta_K)+d\Gamma(\hat\phi,\theta_\ell,\pi-\hat\theta_K)+
d\Gamma(-\hat\phi,\theta_\ell,\hat\theta_K)+d\Gamma(-\hat\phi,\theta_\ell,\pi-\hat\theta_K)$$
where 
\bea \frac{d^4\hat\Gamma}{dq^2\,d\!\cos\hat\theta_K\,d\!\cos\theta_l\,d\hat\phi}=
\frac{9}{32\pi}\left[4 \cos^2\hat\theta_K({\hat F}_L- \beta_\ell^2{\tilde F}_L \cos 2\theta_\ell)+(3 {\hat F}_T+ \beta_\ell^2 {\tilde F}_T \cos 2 \theta_\ell)\sin^2\hat\theta_K \right.&&\nn\\
\left. + 2 {\tilde F}_T (\beta_\ell^2 P_1 \cos 2 \hat\phi \sin^2\hat\theta_K \sin^2\theta_\ell+ 4 \beta_\ell P_2 \cos\theta_\ell \sin^2\hat\theta_K) \right] \frac{d\Gamma_{K*}}{dq^2}  + W_8\quad \eea
with $$W_8=\frac{1}{\pi}\left[{\tilde J}_{1a}^c+{\tilde J}_{2a}^c \cos 2 \theta_\ell+\cos\hat\phi ({\tilde J}_5 + 
2 {\tilde J}_4 \cos\theta_\ell)\sin\hat\theta_K \sin\theta_\ell\right]$$

IX. Identifying $\phi\leftrightarrow-\phi$ when $\phi<0$ and $\theta_K \leftrightarrow \theta_K -\frac{\pi}{2}$ when $\theta_K>\frac{\pi}{2}$ with $\hat \phi \in [0,\pi]$ and $\hat \theta_K \in [0,\pi/2]$ with the combined distribution
$$d\hat\Gamma=d\Gamma(\hat\phi,\theta_\ell,\hat\theta_K)+d\Gamma(\hat\phi,\theta_\ell,\hat\theta_K+\frac{\pi}{2})+
d\Gamma(-\hat\phi,\theta_\ell,\hat\theta_K)+d\Gamma(-\hat\phi,\theta_\ell,\hat\theta_K+\frac{\pi}{2})$$
where 
\bea \frac{d^4\hat\Gamma}{dq^2\,d\!\cos\hat\theta_K\,d\!\cos\theta_l\,d\hat\phi}=
\frac{9}{64\pi}\left[4  {\hat F}_L + 3 {\hat F}_T +\beta_\ell^2 (- 4 {\tilde F}_L + {\tilde F}_T)\cos 2 \theta_\ell \right. &&\nn \\
\left. + 2 {\tilde F}_T (\beta_\ell^2 P_1 \cos 2 \hat\phi \sin^2\theta_\ell + 4\beta_\ell P_2 \cos\theta_\ell) \right] \frac{d\Gamma_{K*}}{dq^2}  + W_9\eea
with \bea W_9&=&\frac{1}{2\pi}\left[2 {\tilde J}_{1a}^c+2{\tilde J}_{2a}^c \cos 2 \theta_\ell+
({\tilde J}_{1b}^c + {\tilde J}_{2b}^c \cos 2 \theta_\ell)(\cos \hat\theta_K-\sin \hat\theta_K) \right.\nn \\ &&\left.+ \cos\hat\phi ({\tilde J}_5+ 2
{\tilde J}_4 \cos\theta_\ell)(\cos\hat\theta_K+\sin\hat\theta_K) \sin\theta_\ell \right]
\eea
Finally, notice that the list is not exhaustive and other foldings are also possible.



\end{document}